\documentclass[aip,jcp,superscriptaddress,amsfonts,graphicx,preprint,numerical]{revtex4-2}
\usepackage[utf8]{inputenc}
\usepackage[T1]{fontenc}
\usepackage{tgtermes}
\usepackage{amsmath, amssymb}
\usepackage{mathtools}
\usepackage{breqn}
\usepackage{graphicx}
\usepackage{float}
\usepackage{bm}
\usepackage{physics}
\usepackage{xcolor}
\usepackage{verbatim}
\usepackage{enumitem}
\usepackage{chemformula}
\usepackage{hyperref}
\usepackage{multirow,xcolor}
\hypersetup{
    colorlinks = true,
    linkcolor = blue,
    linkbordercolor = white,
    citecolor=blue,
    urlcolor=black
}
\usepackage[export]{adjustbox}
\usepackage{subfig,caption}
\makeatletter
\let\cat@comma@active\@empty
\makeatother

\newcommand{\sups}[1]{\textsuperscript{#1}}
\newcommand{\cms}{cm\textsuperscript{3}/s}
\newcommand{\red}[1]{#1}

\begin{abstract}
    Excited atomic nitrogen atoms play an important role in  plasma formation in hypersonic shock-waves, as happens during spacecraft reentry and other high velocity vehicle applications. In this study, we have thoroughly studied collision induced excitation (CIE) associated with two colliding nitrogen atoms in the \ch{N(^4S)}, \ch{N(^2D)} and \ch{N(^2P)} states at collisions energies up to 6 eV, using time-independent scattering calculations to determine cross sections and temperature-dependent rate coefficients. The calculations are based on potential curves and couplings determined in earlier MRCI calculations with large basis sets, and the results are in good agreement with experiment where comparisons are possible. To properly consider the spin-orbit coupling matrix, we have developed a scaling method for treating transitions between different fine-structure components with calculations that only require calculations with two coupled states, and with this we define accurate degeneracy factors for determining cross sections and rate coefficients that include all states. The results indicate that both spin-orbit and derivative coupling effects can play important roles in collisional excitation and quenching, and that although derivative coupling is always much stronger than spin-orbit, there are many transitions where only spin-orbit can contribute.  As part of this, we identify two distinct pathways associated with \ch{N(^2D)} relaxation, including one Auger-like mechanism leading to 2\ch{N(^2D)} that could be important at high temperature.
\end{abstract}

\begin{document}

\title{Modeling of Collision-Induced Excitation and Quenching of Atomic Nitrogen}
\author{Yanze Wu}
\email{wuyanze@northwestern.edu}
\affiliation{Department of Chemistry, Northwestern University, Evanston, Illinois 60208, USA}
\author{Majdi Hochlaf}
\email{majdi.hochlaf@univ-eiffel.fr}
\affiliation{Université Gustave Eiffel, COSYS/IMSE, 5 Bd Descartes 77454, Champs sur Marne, France}
\author{George C. Schatz}
\email{g-schatz@northwestern.edu}
\affiliation{Department of Chemistry, Northwestern University, Evanston, Illinois 60208, USA}
\date{\today}

\maketitle

\section{Introduction}

Gaseous plasma have unique electromagnetic and thermodynamic properties which can play important roles in various physical and chemical processes in extreme conditions, such as in the shock layer of hypersonic vehicles \cite{park1989Nonequilibrium,olynick1995Comparison,boyd2021Analysis}, combustion \cite{miller1989Mechanism} and the ionosphere \cite{schunk2009Ionospheres}. In the upper atmosphere, excited atomic nitrogen, \ch{N(^2D)} and \ch{N(^2P)}, are key species in plasma formation, as their collisions with atomic oxygen or nitrogen are the major pathways for the associative ionization of NO that produces plasma electrons \cite{peterson1998Dissociative,park1993Review,boyd2021Analysis}. Therefore, an understanding of the formation and consumption of excited atomic nitrogen is highly desired. 

Historically, various pathways for the generation or consumption of excited atomic nitrogen have been investigated, including associative ionization \cite{wilson1966Ionization,ringer1979merged,peterson1998Dissociative,park1993Review,yang2023timedependent}, electron impact excitation \cite{park1993Review,ciccarino2019Electronimpact}, and inelastic collisions between species including \ch{N_2}, \ch{O} and \ch{N}, etc \cite{herron1999Evaluated,aiken2024Collisionalradiative,galvao2013Electronic,lu2023Quantum,lu2023Formation}. Among all these mechanisms, one of the obvious processes is the collision of two neutral \ch{N} atoms. To our knowledge, however, there are very few experimental reports of inelastic cross sections for \ch{N + N} collisions \cite{young1975excitation,taghipour19792p3,polak1976,slovetskii1980Mechanisms}, and all of these have only considered the quenching of \ch{N(^2P)} at a single temperature. We therefore believe that a high quality theoretical study would be of significant importance, providing a comprehensive description of the energy transfer kinetics as well as a benchmark for future experiments. Specifically, in this paper, we will focus on two processes and their reverse:
\begin{align}
    \label{reac:n4s_n2d} \tag{R1} \ch{N(^4S) + N <-> N(^2D) + N \\ 
    N(^2D) + N <-> N(^2P) + N  } \label{reac:n2d_n2p} \tag{R2}
\end{align}
where N without any annotation represents the ground state nitrogen, \ch{N(^4S)}. Process \ref{reac:n2d_n2p} has received more attention in the literature, especially in afterglow studies. However, Process \ref{reac:n4s_n2d} is also an important mechanism as it directly produces excited nitrogen \ch{N(^2D)} from the ground state, and this species has been shown to be important to shock-wave ionization in a recent report \cite{aiken2024Collisionalradiative}. 

We will also briefly discuss two other processes: First, the Auger-like quenching of \ch{N(^2P)}:
\begin{align}
    \ch{N(^2P) + N -> N(^2D) + N(^2D)} \label{reac:n2p_n2d2d} \tag{R3}
\end{align}
which is endothermic and is rarely included in kinetic models of plasma formulation. However we find that it may be competitive with Process \ref{reac:n2d_n2p} at high temperature. Second, we consider the direct excitation pathway to \ch{N(^2P)}:
\begin{align}
    \ch{N(^4S) + N <-> N(^2P) + N} \label{reac:n4s_n2p} \tag{R4}
\end{align}
Earlier experimental studies usually reported Process \ref{reac:n4s_n2p} instead of Process \ref{reac:n2d_n2p}, while in later work the products were believed to involve \ch{N(^2D)}. Given this confusion, it would be helpful to consider \ref{reac:n4s_n2p} for clarification.

High quality potential energy curves (PECs) are crucial for quantum dynamical studies. Fortunately, the potential energy curves of \ch{N_2} have been extensively studied in the literature \cite{spelsberg2001Dipoleallowed,hochlaf2010Valence,guberman2012Spectroscopy,little2013initio,ding2023Collision,gelfand2023Nonadiabatic,gelfand2024Nonadiabatic}. In this paper, we adopt the PECs and spin-orbit couplings (SOCs) reported by Hochlaf {\em et al} \cite{hochlaf2010Valence} for all the states except for the $\mathrm{^3\Pi_u}$ states and for the relevant spin-orbit couplings and nonadiabatic (i.e., derivative) couplings (NACs). For these special states, we use recent data from Gelfand {\em et al} \cite{gelfand2023Nonadiabatic}. Although PECs from the two sources were generated using slightly different electronic structure methods, we find them generally compatible, as described in detail in Sec.~\ref{sec:mixing_PEC}. 
Note that in this paper, we adopt the spin-diabatic picture, and by \textit{nonadiabatic couplings} we always refer to the derivative coupling between states with same spin and same symmetry.

\red{In the literature, apart from the spin-orbit couplings and nonadiabatic (derivative) couplings, radiative transitions are often mentioned is contributing to the excited state dynamics of \ch{N_2}, especially in photodissociation processes \cite{hochlaf2010Quintet,ayari2020Statetostate}. As the radiative transitions typically have a much longer timescale (about $10^{-6}$ s) than the lifetime of resonant intermediates (about $10^{-11}$ s based on results presented later), we will not consider these pathways in this paper.}

For the rate calculation, we use a time-independent formalism to solve the Schrodinger equation, which is a widely used approach in atomic scattering calculations \cite{schatz1996Scattering}. For diatomic systems, the most accurate quantum treatment would involve a multidimensional basis with both the internuclear distance $R$ and the (nuclear) rotational angular momentum $J$, however such coupled-channel calculations are computationally expensive and difficult to implement given the many PEC that are involved. To simplify the calculations, we adopt the following approximations: (1) we assume that the rotational angular momentum remains constant, which allows us to perform calculations that involve a few coupled states for each $J$ in parallel; (2) we assume that the spin-orbit couplings are perturbative so we only need to consider one pair of sublevels of each fine structure transition at a time (but there can still be multiple nonadiabatic crossings). We believe that these assumptions should be sufficiently accurate for high energy scattering problems.

%For Reaction \ref{reac:n2d_n2p}, our quenching rate is in consistency with the experimental data reported by Young and Dunn \cite{young1975excitation}, and Taghipour and Brennen \cite{taghipour19792p3}, however still deviate from the value reported from Polak et al \cite{polak1976,slovetskii1980Mechanisms}. For process \ref{reac:n4s_n2d}, no experimental data is found to our knowledge, however our results are comparable with rate of \ch{N(^4S) + N_2 <-> N(^2D) + N_2}.

The paper is organized as follows: In Sec.~\ref{sec:theory}, we present our selection and processing of the PEC, treatment of spin sublevels, and the details of our scattering calculations. In Sec.~\ref{sec:results}, we present our results for Processes \ref{reac:n4s_n2d} and \ref{reac:n2d_n2p}, the discussions about Processes \ref{reac:n2p_n2d2d} and \ref{reac:n4s_n2p} as well as the mixed usage of PECs. In Sec.~\ref{sec:conclusion}, we present our conclusion.

\section{Theory and Computation Details} \label{sec:theory}

\subsection{Preparation of the PEC} \label{sec:pec}

We obtain the potential energy curves (PECs) from two sources: The potential energy curves of the $\mathrm{^3\Pi_u}$ states, their nonadiabatic couplings, and their SOC with other surfaces are obtained from Gelfand {\em et al}~\cite{gelfand2023Nonadiabatic}, while the rest of the PECs and SOCs are obtained from Hochlaf \textit{et al}~\cite{hochlaf2010Valence}. We made this choice because of the availability of PECs and SOCs for all required transitions in the Hochlaf \textit{et al} work, and of the derivative couplings in the Gelfand \textit{et al} paper.
Although the potential curves in both sources are obtained from a complete active space self-consistent field approach (CASSCF) followed by an internally contracted multireference configuration interaction (MRCI) method, their active spaces and basis sets are different, resulting slight differences in PECs. We will further discuss this mixed usage in the Supplementary Information (SI). All states from both sources are spin-diabats. In Fig.~S1, we plot these diabats for a range of energies that is important to this work.

To ensure the correct thermodynamics, at large internuclear distance, the curves are shifted as well as smoothly extrapolated by an exponential function to match their asymptotic values (Table \ref{table:asymp_energy}). There are two different ways to complete this task: 
\begin{enumerate}
    \item Shift the potential globally by some prescribed value, and then smoothly connect the PEC to the asymptotic value (shift-then-extrapolate);
    \item Extrapolate the PEC first, and then shift the whole curve to match the asymptotic values (extrapolate-then-shift). 
\end{enumerate}
Since shifting the PECs individually may change the energy gap, we use strategy \#1 for most of the states. The shift value is determined as 1750 cm\sups{-1} for the $\mathrm{^3\Pi_u}$ states, and 410 cm\sups{-1} for the rest of the states. However, for $\mathrm{1^3\Sigma_g^-,{B'}^3\Sigma_g^-,2^5\Pi_u}$ and $\mathrm{{C''}^5\Pi_u}$, their PECs are attractive at large distance and the asymptotic behavior of the PEC strongly influences the low temperature dynamics, therefore we use strategy \#2. We do not find any significant change of crossing positions brought by such state-wise energy shifting.
Further details of the fitting and regularization are described in the SI. 

\begin{table}[h]
\begin{center}
\begin{tabular}{ c|c } 
 \hline
 Asymptote & Energy/cm$^{-1}$ \\ 
 \hline
 \ch{N(^4S) + N(^4S)} & 78682 \cite{bytautas2005Correlation} \\
 \ch{N(^2D) + N(^4S)} & 97910 \cite{moore1971Natl} \\
 \ch{N(^2P) + N(^4S)} & 107521 \cite{moore1971Natl} \\
 \ch{N(^2D) + N(^2D)} & 117138 \cite{moore1971Natl} \\
 \hline
\end{tabular}
\end{center}
\caption{The asymptotic energy of N + N for the different electronic states we use. These values are respect to the $\nu=0$ vibrational state on the \ch{N_2} ground energy surface ($\mathrm{X^1\Sigma_g^+}$).} \label{table:asymp_energy}
\end{table}

\subsection{Selection Rules and the SOC Scaling}

Due to the high symmetry of the \ch{N_2} molecule, there are a few selection rules for the allowed transitions through either spin-orbit couplings (intersystem crossings) or the nonadiabatic (derivative) couplings (internal conversions). 

For the internal conversion processes, the electronic orbital angular momentum $\Lambda$, spin angular momenta $\Sigma$ and $m_S$, and parity should be the same for the initial and final states. For intersystem crossings, we have $\Delta\Lambda=0/\pm 1$, $\Delta \Sigma=0/\pm 1$, $\Delta m_S=0/\pm 1$, $\Delta\Omega = 0$ and parity is unchanged. Furthermore, for the $\Sigma$ states, the initial and final states must have different reflection symmetry, i.e., a $\Sigma^+$ state can couple to a $\Sigma^-$ state through SOC, but not to another $\Sigma^+$ state.

Since a PEC typically involves multiple sublevels with different $m_S$, the spin-orbit coupling between two PECs comes as a matrix with size $m_{S,1} \times m_{S,2}$ instead of a scalar. For the \ch{N_2} system, most of the active states are triplets or quintets, therefore more than 10 sublevels (considering the orbital degeneracy) are typically involved in the calculation on a single intersystem crossing. However, in the golden rule limit (as in our case), we can calculate the transition probability using only two sublevels, and scale up the resulting transition probability by taking advantage of the SOC matrix's structure.

To see how such scaling works, recall the general form of the SOC matrix
\begin{align}
    V_{SO} = V_{SO}^xS_x + V_{SO}^yS_y + V_{SO}^zS_z
\end{align}
where $V_{SO}^{x/y/z}$ are pure-imaginary scalars and $S_{x/y/z}$ are the coefficient spin matrices. For a diatomic system, when the proper symmetry convention is chosen, only one of $V_{SO}^x,V_{SO}^y$ or $V_{SO}^z$ is nonzero. Therefore we can define the scaling factor that connects the state-to-state transition probability to the sum over degenerate states as
\begin{align}
    \zeta = \frac{\norm{S_\alpha}^2}{\abs{\mel{m_S^{[1]}}{S_\alpha}{m_S^{[2]}} }^2}
\end{align}
where $\alpha=x,y$ or $z$ depending on the states' symmetry, the $\norm{S_\alpha}$ represents the Frobenius norm of the matrix, and $m_S^{[1]}$ and $m_S^{[2]}$ are the sublevels used in the calculation (typically ones with the maximum $m_S$). For example, in a triplet-quintet crossing with only $V_{SO}^y$, the matrix looks like (where the basis are ordered by their $m_S$)
\begin{align}
    V_{SO} = \begin{bmatrix} \sqrt{6} & 0 & 1 & 0 & 0 \\ 0 & \sqrt{3} & 0 & \sqrt{3} & 0 \\ 0 & 0 & 1 & 0 & \sqrt{6} \end{bmatrix}V_{SO}^y \label{eq:zeta}
\end{align}
If we use the top-left block in our scattering calculation (i.e., a coupling associated with $\sqrt{6}V_{SO}^y$), then the scaling factor is $\zeta = (6+3+1)\times 2 / 6 = 10/3$.

\subsection{The Scattering Calculation}

The channel-specific transition probabilities are obtained from time-independent quantum scattering calculations. For the \ch{N_2} system, there are 4 PECs belonging to the \ch{N(^4S) + N(^4S)} asymptote, 11 PECs belonging to the \ch{N(^4S) + N(^2D)} asymptote and 8 PECs to the \ch{N(^4S) + N(^2P)} asymptote. While including all the relevant potential curves in a single calculation is possible, it greatly complicates the calculation and makes the interpretation difficult. Instead, we take advantage of the perturbative nature of the spin-orbit couplings (note that we still treat the nonadiabatic (derivative) couplings as non-perturbative), and include only a few states (a ``calculation group'') in a single calculation. There are several ways to construct a calculation group:
\begin{enumerate}
    \item Direct intersystem crossings. We start with two states: one PEC from the reactant asymptote, and the other from the product asymptote. They must have at least one allowed intersystem crossing to be considered. If any of the two PEC couple to other PECs via NAC, all the coupled states are included. 
    \item Direct nonadiabatic couplings. Again we start with the two nonadiabatically coupled PECs, one from the reactant asymptote and the other from the product asymptote. Any other PECs coupled with these two PECs via NAC are included.
    \item Indirect crossings. Since the nonadiabatic couplings are treated non-perturbatively, it's possible that potential curve A has intersystem crossing with curve B, and B to C through NACs. In the end, there are a chain of coupled PECs. Note that \red{as we treated SOC as perturbative, } in such a chain, only one intersystem crossing is allowed at most. \red{Indirect process involving multiple intersystem crossings, e.g., those involving $^1\Sigma$ and $^1\Pi$ states, should be considered if higher accuracy is desired in the future.}
    
\end{enumerate}
For the Process \ref{reac:n2d_n2p}, we can pick $\mathrm{1^5\Sigma_u^+}$ (with asymptote \ch{N(^2D) + N(^4S)}) and $\mathrm{2^3\Pi_u}$ (with asymptote \ch{N(^2P) + N(^4S)}) in the group, and then include all the other relevant $\mathrm{^3\Pi_u}$ states ($\mathrm{1^3\Pi_u}$, $\mathrm{3^3\Pi_u}$, etc.) in the group, as they are nonadiabatically coupled to $\mathrm{2^3\Pi_u}$.

After the calculation groups are determined, we use the following Hamiltonian in our quantum scattering calculation:
\begin{align}
    H = T_{nu} + \frac{J(J+1)}{2MR^2} + \sum_{m,n}{\left(V_{mn}(R) - i\frac{P_{nu}}{M}d_{mn}(R) \right)\ket{m(R)}\bra{n(R)}}
\end{align}
where $T_{nu}$ is the nuclear kinetic energy operator, $P_{nu}$ is the nuclear momentum operator, $M$ is the reduced mass, $J$ is the rotational angular momentum, $V_{mn}$ is the spin-orbit coupling, $d_{mn}$ is the derivative coupling with respect to the radial coordinate, and $m,n$ iterate through all states in the calculation group. 

Note that we do not include any coupling terms between different $J$ channels, so the problem remains 1D for each asymptotic state. This approximation is commonly used in semiclassical calculations \cite{bieniek1977Uniform}, and despite missing electronic Coriolis effects, it should be relatively accurate when $J_{nuclei} \gg J_{electron}$. For our current problem, $J_{nuclei}$ is typically 40-200, and $J_{electron} \le 3$, therefore we believe the assumption is accurate. Nevertheless, future calculations with coupled $J$ channels should be carried out to confirm this assumption.

The scattering calculation is carried out by solving the Schrodinger equation with plane-wave boundary conditions. Suppose the reactant is incident from state $n_{\text{inc}}$ with kinetic energy $E$, then the equation to be solved is
\begin{align}
    (E_{tot} - H)\left(\sum_n{\psi_n(R)\ket{n(R)}} + \sum_{\text{allowed }n}{r_n\ket{\psi^{\text{refl}}_n}} + \ket{\psi^{\text{inc}}}\right) = 0 \label{eq:scattering}
\end{align}
where $E_{tot} = E + V_{n_{\text{inc}}n_{\text{inc}}}(+\infty)$ is the total energy, and $\psi_n(R)$ and $r_n$ are the wavefunction and reflection coefficients to be determined. The $\ket{\psi^{\text{inc}}}$ and $\ket{\psi_n^{\text{refl}}}$ are the plane-wave boundary basis, defined as
\begin{align}
    \ket{\psi^{\text{inc}}} &= \sum_{R}{e^{-ik_{n^{\text{inc}}}R}\ket{n_{\text{inc}}(R)}} \label{eq:psi_inc} \\
    \ket{\psi_n^{\text{refl}}} &= \sum_{R}{e^{ik_{n}R}\ket{n(R)}} \label{eq:psi_refl}
\end{align}
Here in Eqs.~\eqref{eq:psi_inc} and \eqref{eq:psi_refl}, $k_n = \sqrt{E_{tot} - V_{nn}(R_{max}) - J(J+1)/(2MR_{max}^2)}$ is the wavevector. Only states with a real-valued $k_n$ are included in the calculation. The transition probability for the outgoing state $n_{\text{out}}$ is given by 
\begin{align}
    P_{n_{\text{out}}} = \frac{\abs{r_{n_{\text{out}}}}^2 k_n}{k_{n_{\text{inc}}}}
\end{align}
In the scattering calculations, both the kinetic energy operator and momentum operator are evaluated by 9-point finite difference approximations. For calculation groups with intersystem crossings only, we use a Cartesian grid of 1600 elements, spanned as $[R_{min},R_{max}]$ = [1 au, 24 au]. If nonadiabatic couplings are involved, we use a grid of 3000 elements spanned as [1 au, 16 au].

After the channel-specific transition probability is obtained, we evaluate the cross section using \cite{sun1992Collisional}
\begin{align}
    \sigma_{m\to n}(E) = g_{m\to n}\frac{2\pi}{k^2_m}\sum_{J}{g_{nu}(J)(2J+1)P_{m\to n}(E)}
\end{align}
Here, $g_{nu}(J)$ accounts for the fact that the two nitrogen-14 nuclei are identical with spin 1, and equals to 2/3 for even $J$ and 1/3 for odd $J$ \cite{ochkin2009Spectroscopy}. The parameter $g_{m\to n}$ is the effective electronic degeneracy. For a calculation group with intersystem crossing, $g_{m\to n}$ is defined by
\begin{align}
    g_{m\to n}^{\text{SOC}} = \frac{\zeta(2-\delta_{\Lambda_1,0}\delta_{\Lambda_2,0}) }{N_p(2S_1+1)(2L_1+1)(2S_2+1)(2L_2+1)} \label{eq:degeneracy}
\end{align}
where $S_1,S_2$ and $L_1,L_2$ are the initial spin and orbital angular momenta of the two N atoms, $\Lambda_1,\Lambda_2$ and $\Sigma_1,\Sigma_2$ are the electronic orbital and spin angular momenta of the two curves involved in the intersystem crossing, and $\zeta$ is the scaling factor defined by Eq.~\eqref{eq:zeta}. $N_p$ is a parity factor, which equals to 2 if the two atoms are initially in same electronic state, and 1 otherwise. If all states are coupled through NACs, $g_{m\to n}$ is defined by
\begin{align}
      g_{m\to n}^{\text{NAC}} = \frac{(2-\delta_{\Lambda_1,0})(2\Sigma_1+1)}{N_p(2S_1+1)(2L_1+1)(2S_2+1)(2L_2+1)}
\end{align}

Finally, the temperature-dependent rate coefficient is given by averaging the cross sections over a Maxwell-Boltzmann distribution of kinetic energies:\cite{lu2023Quantum}
\begin{align}
    k(T) = \sqrt{\frac{8}{\pi M(k_BT)^3}}\sum_{m,n}{\int_0^\infty{e^{-E/k_B T}\sigma_{m\to n}(E)EdE }}
\end{align}
The rate coefficient is then fitted according to the generalized Arrhenius formula:
\begin{align} \label{eq:arrhenius}
    \log{k} = -\frac{E_a}{T} + \eta \log{T} + A
\end{align}

\section{Results and Discussion} \label{sec:results}

\subsection{\ch{N + N <-> N(^2D) + N}}

\begin{figure}[h]
    \centering
    \includegraphics[width=0.8\textwidth]{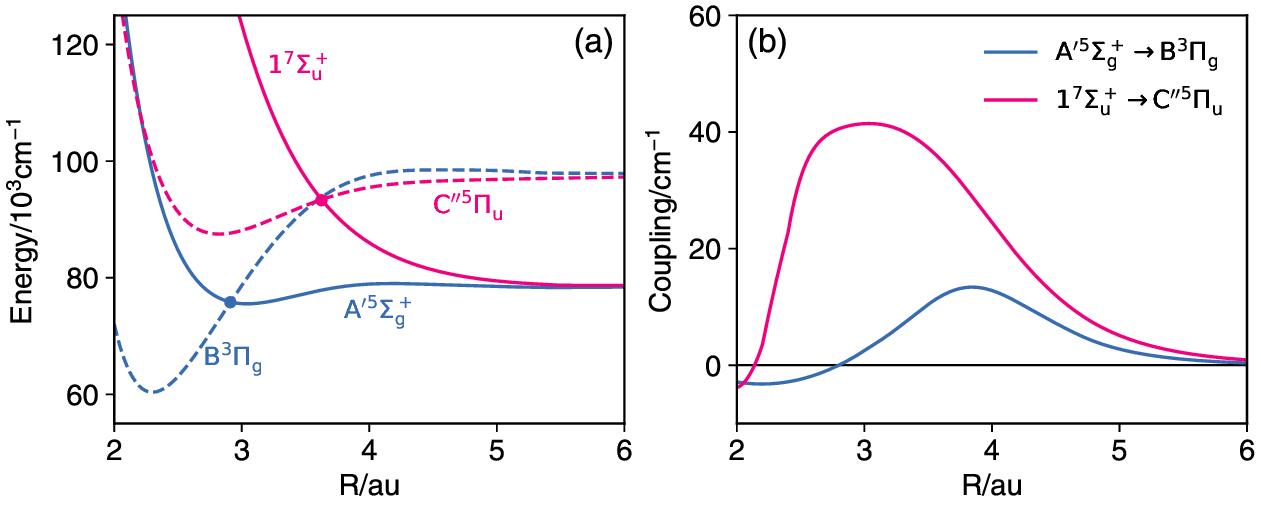}
    \caption{(a) Potential curves relevant to \ref{reac:n4s_n2d}. There are two pairs of crossings that are relevant: $\mathrm{1^7\Sigma_u^+-{C''}^5\Pi_u}$ and $\mathrm{{A'}^5\Sigma_g^+-B^3\Pi_g}$. (b) Spin-orbit couplings for the crossings in (a).}
    \label{fig:4s2d}
\end{figure}

A total of two crossings, $\mathrm{1^7\Sigma_u^+-{C''}^5\Pi_u}$ and $\mathrm{{A'}^5\Sigma_g^+-B^3\Pi_g}$, are relevant to the Process \ref{reac:n4s_n2d}. These potential curves and crossing points are plotted in Fig.~\ref{fig:4s2d}a. As shown in Fig.~\ref{fig:4s2d}b, the SOC between $\mathrm{{A'}^5\Sigma_g^+}$ and $\mathrm{B^3\Pi_g}$ has only a tiny value (1.2 cm\sups{-1}) at the crossing point because of a sign change nearby. As a consequence, the transition between $\mathrm{{C''}^5\Pi_u}$ and $\mathrm{1^7\Sigma_u^+}$ dominates both the excitation and quenching processes, with a cross section that is three orders of magnitude larger than for the $\mathrm{{A'}^5\Sigma_g^+\leftrightarrow B^3\Pi_g}$ transition (not shown).

\begin{figure}[H]
    \centering
    \includegraphics[width=0.8\textwidth]{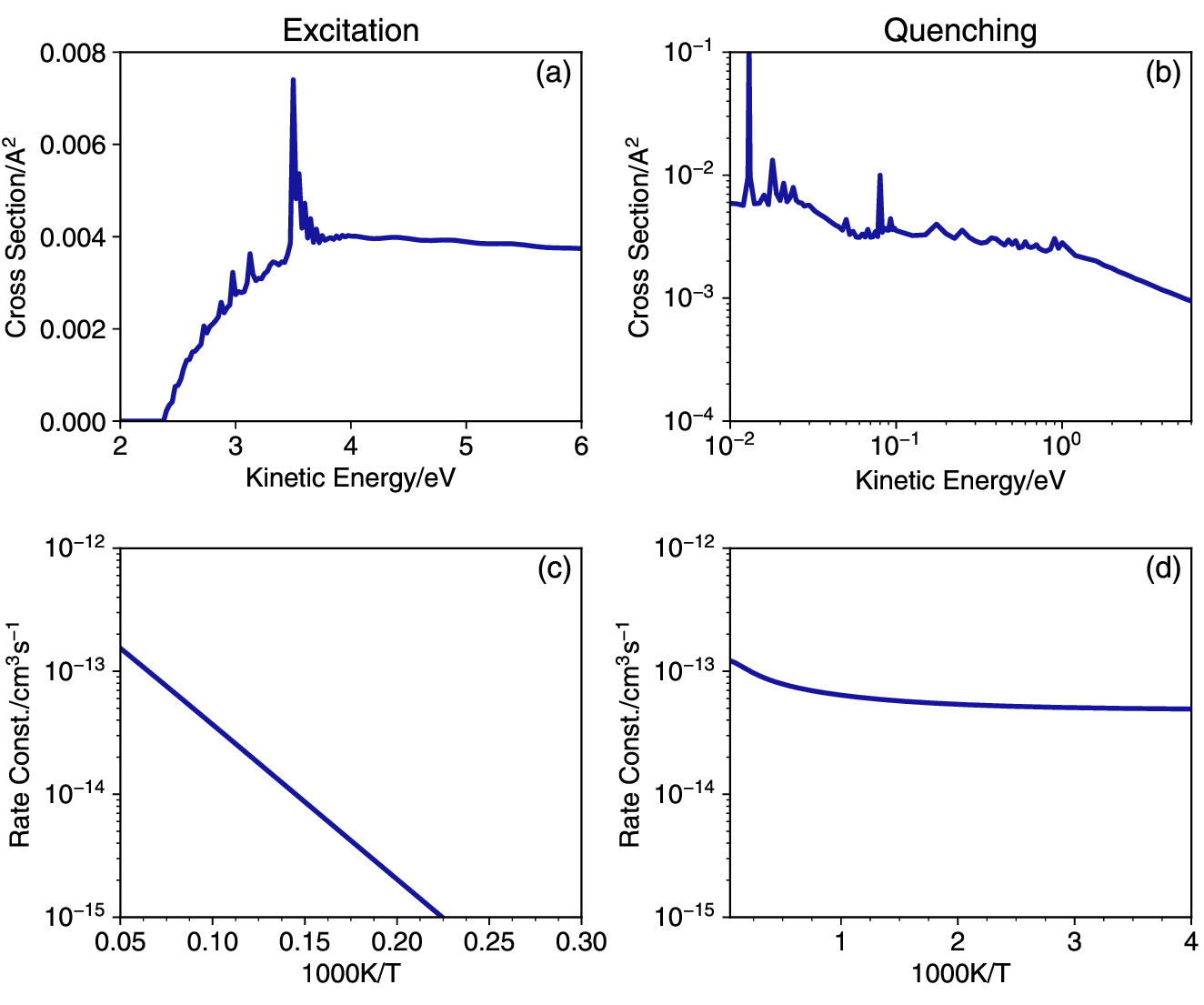}
    \caption{(a)(b): Excitation and quenching cross sections for \ref{reac:n4s_n2d}. (c)(d): Arrhenius plot of the corresponding rate coefficients.}
    \label{fig:4s2d_rate}
\end{figure}

Cross sections for the $\mathrm{1^7\Sigma_u^+\to{C''}^5\Pi_u}$ transition and its reverse are shown in Fig.~\ref{fig:4s2d_rate}a and Fig.~\ref{fig:4s2d_rate}b. The cross section for excitation is nonzero when the kinetic energy is above 2.38 eV, consistent with the asymptotic energy gap. \red{There is a peak at 3.5 eV due to the formation of resonance states on the $\mathrm{{C''}^5\Pi_u}$ surface.}
For quenching, due to the absence of a potential energy barrier, the cross section for $\mathrm{{C''}^5\Pi_u}\to\mathrm{1^7\Sigma_u^+}$ decreases almost monotonically except a few resonance peaks \red{on the $\mathrm{{C''}^5\Pi_u}$ surface}.

The temperature-dependent rate coefficients for both forward and backward processes are shown in Fig.~\ref{fig:4s2d_rate}c and Fig.~\ref{fig:4s2d_rate}d, and the values are provided in the SI. The best fit parameters using Eq.~\eqref{eq:arrhenius} are presented in Table \ref{table:rate_4s2d}. The excitation process has a rate coefficient of $3.7\times 10^{-14}$ \cms{} at 10000 K, with an activation energy close to the 2.38 eV energy barrier. The quenching rate coefficient at 300 K is $5.0\times 10^{-14}$ \cms, with an almost negligible energy barrier. These values are similar to the rate coefficients for \ch{N + N_2 <-> N(^2D) + N_2} ($5\times 10^{-14}$ \cms{} for excitation at 10000 K, and $1\times 10^{-14}$ \cms{} for quenching at 300 K) as recently reported by Lu et al \cite{lu2023Quantum}, indicating that the two processes may share a similar mechanism.
\begin{table}[H]
    \centering
    \begin{tabular}{c|c|c|c}
        \hline
        Process & $A$ (\cms) & $\eta$ & $E_a$ (K) \\
        \hline
         Excitation & $6.44 \times 10^{-13}$ & 0 & $2.88\times 10^4$ \\
         Quenching & $1.36\times 10^{-14}$ & 0.23 & 31.2 \\
         \hline
    \end{tabular}
    \caption{Fitted parameters of the Arrhenius formula for Process \ref{reac:n4s_n2d}.}
    \label{table:rate_4s2d}
\end{table}

\subsection{\ch{N(^2D) + N <-> N(^2P) + N}} \label{sec:results_2d2p}

\begin{figure}[h]
    \centering
    \includegraphics[width=0.8\textwidth]{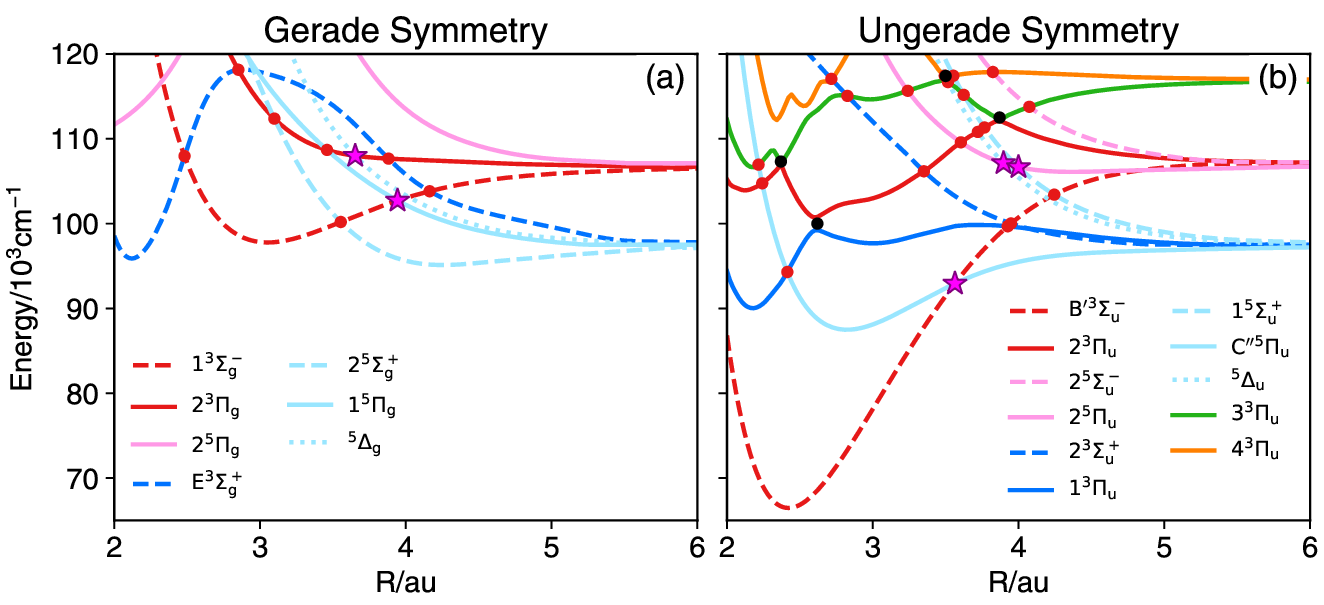}
    \caption{Curves and crossings relevant to \ref{reac:n2d_n2p}. The curves are categorized according to their parity: (a) gerade (b) ungerade. Red dots mark the less important intersystem crossings, while pink stars mark the ones \red{that contribute to the most of the rates}. Black dots mark the nonadiabatic crossings. }
    \label{fig:2d2p_surf}
\end{figure}

As shown in Fig.~\ref{fig:2d2p_surf}, a total of 18 potential curves (which form 20 calculation groups) are found for \ref{reac:n2d_n2p}, indicating that this process is substantially more complex than \ch{N(^4S) <-> N(^2D)}. Details of the calculation groups and degeneracy factors are listed in Table S2 in the SI.

\begin{figure}[h]
    \centering
    \includegraphics[width=0.9\textwidth]{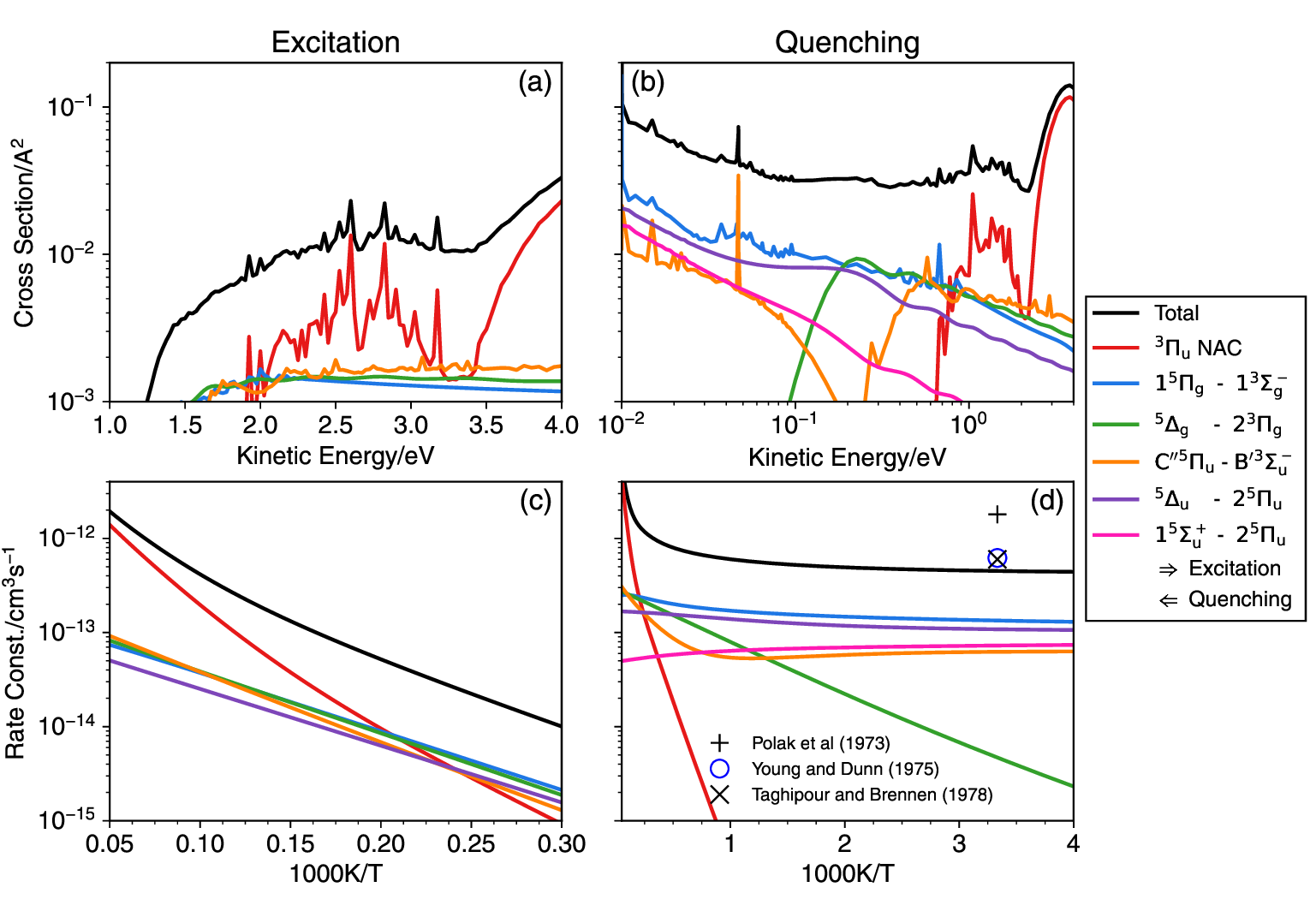}
    \caption{(a)(b): Excitation and quenching cross sections for \ref{reac:n2d_n2p}. The contribution of several important pathways are also included. The $\mathrm{2^5\Pi_u\to 1^5\Sigma_u^+}$ pathway is not shown for excitation since its cross section is negligible. (c)(d): Arrhenius plots of the corresponding rate coefficients.}
    \label{fig:2d2p_rate}
\end{figure}

Cross sections for excitation and quenching are shown in Fig.~\ref{fig:2d2p_rate}a and \ref{fig:2d2p_rate}b, respectively, where we also plot the cross sections for a few important pathways (marked in Fig.~\ref{fig:2d2p_surf}): $\mathrm{1^3\Sigma_g^-\to 1^5\Pi_g}$, $\mathrm{2^5\Pi_u\to {^5}\Delta_u}$, $\mathrm{2^5\Pi_u\to 1^5\Sigma_u^+}$, $\mathrm{{B'}^3\Sigma_u^-\to {C''}^5\Pi_u}$, $\mathrm{2^3\Pi_g\to {^5}\Delta_g}$ and the nonadiabatic transitions associated with the $\mathrm{^3\Pi_u}$ states. These pathways typically contribute more than 80\% to the rate coefficient (see SI for the detailed values). \red{For the quenching process at low collisional energy, there are a few resonance peaks, e.g., the 0.047 eV peak on $\mathrm{{B'}^3\Sigma_u^-\to {C''}^5\Pi_u}$ due to the formation of the resonance states on the $\mathrm{{B'}^3\Sigma_u^-}$ surface. We further examine the 0.047 eV resonance peak and find its full width at half maximum (FWHM) is about $6\times 10^{-5}$ eV, corresponding to a lifetime of 10 ps.}

\begin{figure}[h]
    \centering
    \includegraphics[width=0.6\textwidth]{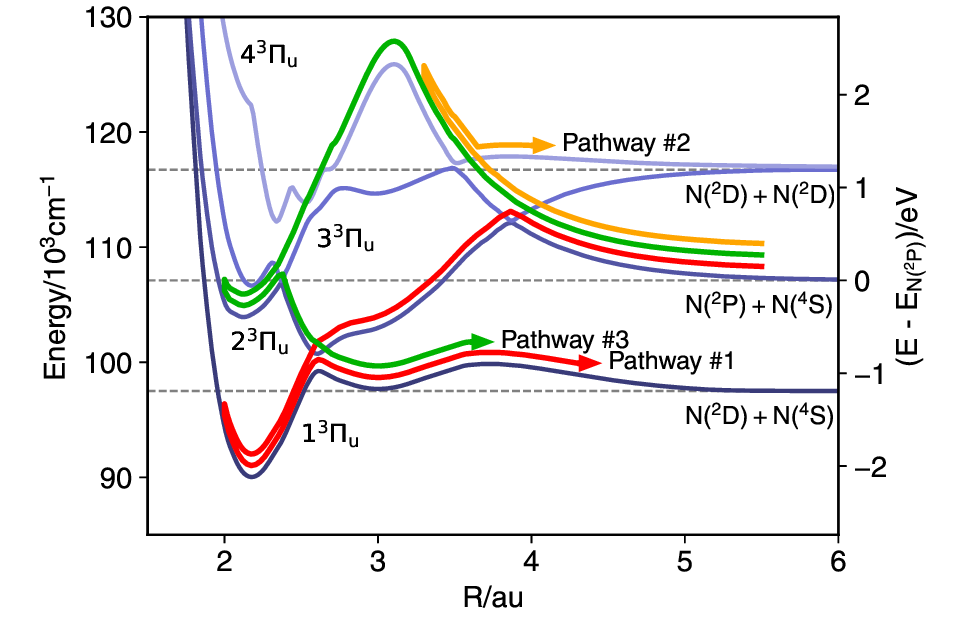}
    \caption{The $\mathrm{^3\Pi_u}$ states and the main nonadiabatic transition pathways for different collisional energies that lead to quenching. For lower collisional energy, the population follows Pathway \#1 ($\mathrm{2^3\Pi_u \to 1^3\Pi_u}$); for higher energy, some of the population goes through Pathway \#2 ($\mathrm{2^3\Pi_u \to 3^3\Pi_u \to 4^3\Pi_u}$) and ends up giving two \ch{N(^2D)}. For even higher collisional energy, the population goes through the barrier on $\mathrm{4^3\Pi_u}$ and goes down back to \ch{N(^2D) + N(^4S)}. Note that there can be additional pathways at higher collisional energies if the higher lying $\mathrm{^3\Pi_u}$ states are included. }
    \label{fig:3piu}
\end{figure}

At a low collisional energy, all available pathways for both quenching and excitation are mediated by SOC, as the nonadiabatic transitions between $\mathrm{^3\Pi_u}$ states are forbidden by a barrier (0.6 eV for quenching, 1.8 eV for excitation) on the $\mathrm{3^3\Pi_u}$ state.
Once the collisional energy exceeds the barrier, nonadiabatic transitions involving derivative coupling quickly dominate the scattering mechanism, leading to Pathway \#1 in Fig.~\ref{fig:3piu}. As the kinetic energy further increases, the nonadiabatic cross sections begin to drop because of the opening of the \ch{N(^2D) + N(^2D)} channel (Pathway \#2). This Auger-like conversion is likely to be an important mechanism for generation of \ch{N(^2D)} at high temperature, as further discussed in Sec.~\ref{sec:n2p_n2d2d}. 
The nonadiabatic cross section is also important when the collisional energy exceeds the barrier on the $\mathrm{4^3\Pi_u}$ state (about 2.5 eV for quenching, or 3.5 eV for excitation). In this energy range, the scattering process involves a roundabout mechanism, going up through several nonadiabatic crossings to the $\mathrm{4^3\Pi_u}$ state before coming down to the target state (Pathway \#3).
Note that since we only consider the four lowest $\mathrm{^3\Pi_u}$ states, our calculated cross section does not significantly drop at the high energy limit. In reality, we expect a lower and more oscillatory cross section at higher collisional energy, as the collisions access higher adiabats (especially $\mathrm{5^3\Pi_u}$ and $\mathrm{6^3\Pi_u}$) through nonadiabatic transitions. As a consequence, we also expect the actual value of rate coefficients at $T >$ 20000 K to be slightly smaller than in our calculation.

The relative importance of SOC/NAC pathways at different kinetic energies is also seen in the temperature dependence of the rate coefficients. As shown in Fig.~\ref{fig:2d2p_rate}c and \ref{fig:2d2p_rate}d, the Arrhenius plots have different slopes in different temperature regions. The change in pathway is estimated to occur near 10000 K for both quenching and excitation. In order to describe this multistage behavior of the $k-T$ relationship, we fit the rate coefficients of the nonadiabatic and SOC pathways separately, see Table \ref{table:rate_2d2p}. The overall rate coefficient is the sum of the two pathways.
\begin{table}[h]
    \centering
    \begin{tabular}{c|c|c|c}
        \hline
        Process & $A$ (\cms) & $\eta$ & $E_a$ (K) \\
        \hline
         Excitation (NAC) & $7.52\times 10^{-20}$ & 1.78 & $1.68\times 10^4$ \\
         Excitation (SOC) & $3.59\times 10^{-14}$ & 0.35 & $1.39\times 10^4$ \\
         Quenching (NAC) & $3.91\times 10^{-18}$ & 1.43  & $4.95\times 10^3$ \\
         Quenching (SOC) & $3.05\times 10^{-14}$ & 0.42 & -89.9 \\
         \hline
    \end{tabular}
    \caption{Parameters of the Arrhenius formula for Process \ref{reac:n2d_n2p}. The overall rate coefficient is the sum of the NAC and SOC pathways.}
    \label{table:rate_2d2p}
\end{table}

At 300 K, our calculated value of the quenching rate coefficient is $4.5\times 10^{-13}$ \cms, which is only slightly lower than the experimental value of $6.2\times 10^{-13}$ \cms{} from Young and Dunn \cite{young1975excitation} and $6\times 10^{-13}$ \cms{} from Taghibour and Brennen \cite{taghipour19792p3}. Given uncertainties in the experimental measurements that were estimated as at least 20\% as well as errors due to asymptotic PECs in the electronic structure calculations, we believe that our result is consistent with these experiments. Interestingly, there is another widely used value in the literature from Polak et al, who report a result of $1.8\times 10^{-12}$ \cms{} \cite{polak1976,slovetskii1980Mechanisms}. Even though this number is unlikely to be the room temperature value, it could still be representative in the shock-wave region where the temperature is above 3000 K and the quenching rate coefficient is at the order of $10^{-12}$ \cms.

\subsection{\ch{N(^2P) + N <-> N(^2D) + N(^2D)}} \label{sec:n2p_n2d2d}

As mentioned in Sec.~\ref{sec:results_2d2p}, when the kinetic energy is sufficient, the NAC dynamics may lead to the production of two \ch{N(^2D)} atoms. In this section we calculate the cross section and rate coefficients for this process.

\begin{figure}[H]
    \centering
    \includegraphics[width=0.9\textwidth]{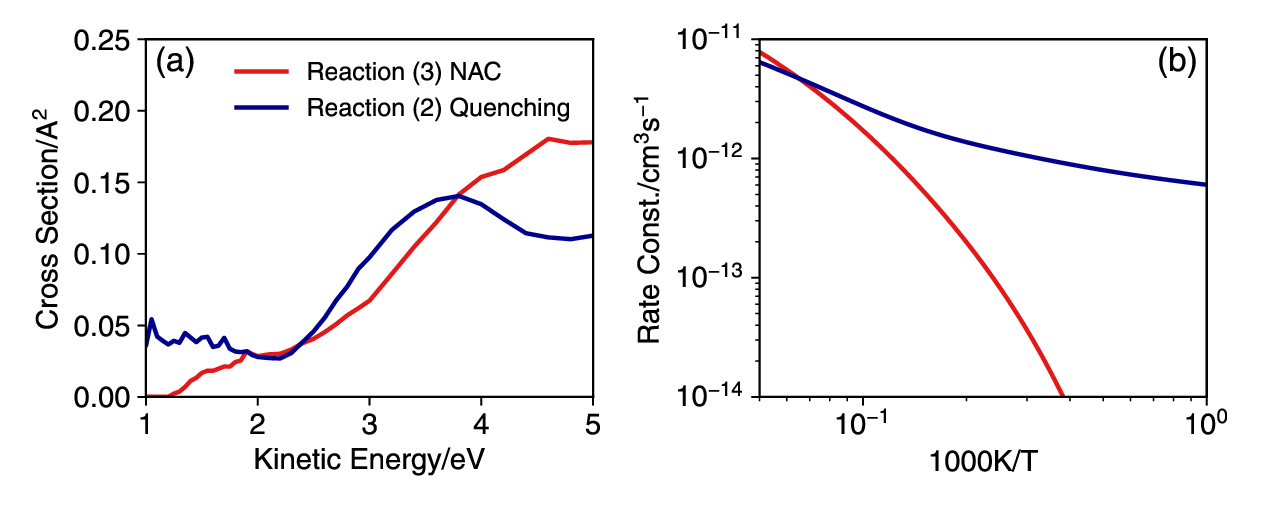}
    \caption{(a) Reactive cross section and (b) rate coefficients for Process \ref{reac:n2p_n2d2d} associated with the $\mathrm{^3\Pi_u}$ nonadiabatic pathway, where we also plot the quenching Process \ref{reac:n2d_n2p} for comparison. At high kinetic energy (> 3.8 eV) or high temperature (> 15000 K),  \ref{reac:n2p_n2d2d} is preferred to the quenching pathway.} 
    \label{fig:n2p_n2d2d}
\end{figure}

As shown in Fig.~\ref{fig:n2p_n2d2d}a, the production of two \ch{N(^2D)} (Process \ref{reac:n2p_n2d2d}) is only preferred when the collisional energy is greater than 3.8 eV. Otherwise, the direct quenching (Process \ref{reac:n2d_n2p}) is preferred. The rate coefficients for both processes are compared in Fig.~\ref{fig:n2p_n2d2d}b, where we see that \ref{reac:n2p_n2d2d} becomes faster than the quenching process \ref{reac:n2d_n2p} at $T >$ 15000 K. These results suggest that  \ref{reac:n2p_n2d2d} is important when the system temperature is relatively high.
Note that the results here are still qualitative, as we have neglected many factors, such as the inclusion of SOC pathways, and the inclusion of higher lying $\mathrm{^3\Pi_u}$ states, therefore these values should only be understood as a rough estimation. If the high-temperature quenching of \ch{N(^2P)} is found to be important enough, a comprehensive study of \ref{reac:n2p_n2d2d} will be necessary.

\subsection{\ch{N(^4S) + N <-> N(^2P) + N}} \label{sec:n4s_n2p}

Finally, it is interesting to consider the possibility of quenching involving \ch{N(^2P)} that leads to both atoms in the ground state \ch{N(^4S)}, as some of the experimental reports originally considered \ch{N(^4S)} as the quenching product starting from \ch{N(^2P)} \cite{slovetskii1980Mechanisms,young1975excitation}. However, after a detailed examination of the potential curves, we find \ch{N(^4S)} is extremely unlikely as a quenching product compared to \ch{N(^2D)}: The least energetic direct pathway for \ch{N(^2P) -> N(^4S)} is $\mathrm{2^5\Pi_u \to 1^7\Sigma_u^+}$, which has a barrier about 1.5 eV. There is also an indirect pathway $\mathrm{2^3\Pi_u \to 1^3\Pi_u \to A^3\Sigma_u^+}$, requiring a barrier of 1.2 eV at minimum. Both of the barriers are sufficiently high to prohibit any processes at low temperature (below 10000 K). At higher temperature, although the quenching process is activated, we expect the majority of the population undergoing nonadiabatic transition to \ch{N(^2D)} instead of going directly to \ch{N(^4S)}. Since the same barrier applies to the reverse process \ch{N(^4S) -> N(^2P)}, the population going directly from \ch{N(^4S)} to \ch{N(^2P)} should also be relatively small compared with other pathways such as \ch{N(^4S) -> N(^2D)}.

\subsection{Compatibility of PECs from the Two Sources} \label{sec:mixing_PEC}

As mentioned in Sec.~\ref{sec:pec}, for the calculation of the \ch{N(^2D) <-> N(^2P)} transition, the $\mathrm{^3\Pi_u}$ surfaces and corresponding couplings are coming from Ref.~\cite{gelfand2023Nonadiabatic} instead of Ref.~\cite{hochlaf2010Valence}. In this section, we provide a comparison of the electronic structure in the two sources.

The potential curves from both sources are calculated based on a complete active space self-consistent field approach (CASSCF) followed by an internally contracted multireference configuration interaction (MRCI) method, however they use different active space definitions and different basis sets. A detailed comparison of the methods from the two sources is presented in Table \ref{table:qm_compare}.

\begin{table}[h]
\begin{center}
\begin{tabular}{ c|p{.3\textwidth}|p{.3\textwidth} } 
 \hline
 Source & Hochlaf et al \cite{hochlaf2010Valence} & Gelfand et al \cite{gelfand2023Nonadiabatic} \\ 
 \hline
Method & \multicolumn{2}{c}{CASSCF/MRCI} \\
 \hline
 Active space & 12 (Valence orbitals + 1$\sigma_g$ + 1$\pi_g$) & 17 ($4\sigma_u,3\sigma_g,4\pi_u,4\pi_g,2\delta_g$) \\
 \hline
 Basis sets & aug-cc-pVQZ + 3s and 2p diffuse GTOs & d-aug-cc-pVQZ + bond-centered (s,p) diffuse functions \\
 \hline
\end{tabular}
\end{center}
\caption{Comparison between the electronic structure methods used in the two sources.} \label{table:qm_compare}
\end{table}

\begin{figure}[h]
    \centering
    \includegraphics[width=0.9\textwidth]{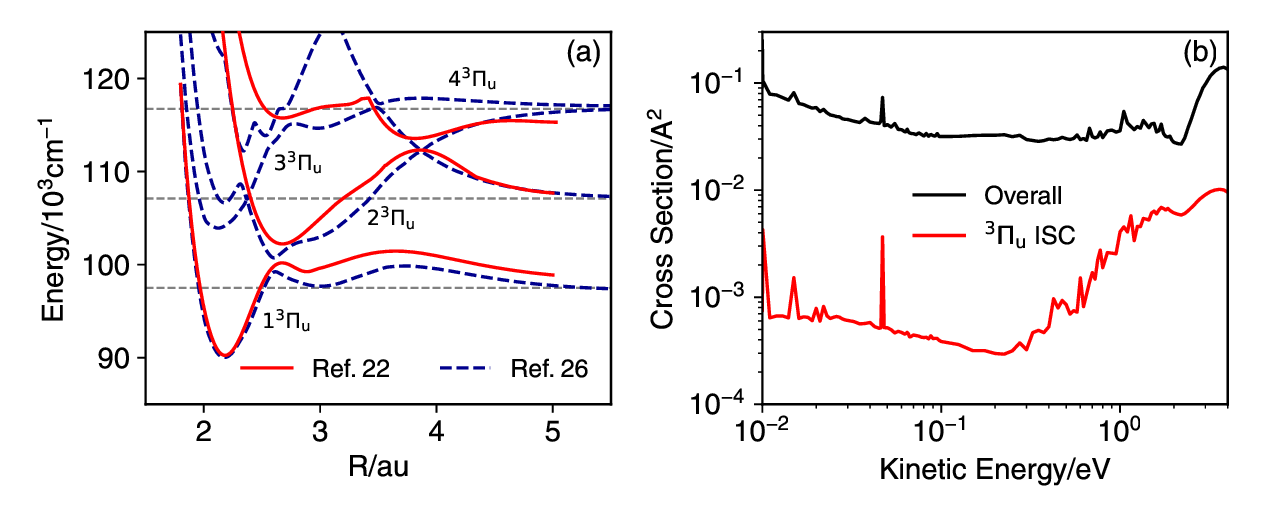}
    \caption{(a) Comparison between the $\mathrm{^3\Pi_u}$ states from Hochlaf et al (2010) \cite{hochlaf2010Valence} (red) and Gelfand et al (2023) \cite{gelfand2023Nonadiabatic} (blue, shifted up by 1340 $\text{cm}^{-1}$ to match the asymptotic energy). The potential curves from the two sources are comparable in general, but have some notable differences, including the energy gap between $\mathrm{2^3\Pi_u}$ and $\mathrm{3^3\Pi_u}$, the barrier on $\mathrm{1^3\Pi_u}$ and the inclusion of other lower-lying states for $R < 3$ au. (b) The cross sections for all SOC pathways involving $\mathrm{^3\Pi_u}$ states, compared with the overall cross section. \red{For most of the energy range, the SOC pathways of $\mathrm{^3\Pi_u}$ states are negligibly small, suggesting changing the PEC sources will not affect the overall cross section through the SOC pathways.} } 
    \label{fig:compare_PEC}
\end{figure}

As shown in Fig.~\ref{fig:compare_PEC}a, the $\mathrm{^3\Pi_u}$ potential curves obtained from the two sources are generally similar, however there are a few noticeable differences: (1) The PECs from Hochlaf et al have a higher energy at $R=3-4$ au, and the barrier on $\mathrm{C^3\Pi_u}$ (or $\mathrm{1^3\Pi_u}$ in Gelfand et al paper) is higher. (2) At short internuclear distance, the PEC from Hochlaf et al doesn't show the avoided crossings that are clearly seen in the Gelfand et al results. In fact, in Ref.~\cite{hochlaf2010Valence} the two lowest PECs are identified by their symmetry ($\mathrm{C^3\Pi_u,{C'}^3\Pi_u}$) suggesting that these are diabats \red{instead of adiabats} \cite{ndome2008Sign}. Although this diabaticity enables computationally simpler scattering calculations, they miss the coupling to the intruder states around $R=2.5$ au \cite{lewis2008coupledchannel} (notice the several nearby avoided crossings at $105-115\times 10^3$ cm\sups{-1} on the PECs of Gelfand et al). These crossings are crucial in describing the high energy (>3 eV) behavior of \ch{N(^2P) -> N(^2D)} transitions.

For low to moderate energy calculations, however, we believe that these differences do not significantly alter the results. First, the barrier on $\mathrm{C^3\Pi_u}$ is not important, as it is much smaller than the \ch{N(^2P) - N(^2D)} energy gap. Second, both sources give a similar energy barrier on $\mathrm{3^3\Pi_u}$, with a difference less than 200 $\text{cm}^{-1}$, therefore the dynamics of Process \ref{reac:n2d_n2p} should be little affected. \red{In fact, the extra active space from Gelfand et al mainly accounts for the Rydberg states, therefore the valence $\mathrm{^3\Pi_u}$ surfaces are indeed expected to be similar to Hochlaf et al results.} Third, apart from the NAC pathways, \red{the states that cross with $\mathrm{^3\Pi_u}$ through SOCs remain the same for the PECs from both sources.} % George: I try to say that the same states cross with 3piu through SOC, given whatever the source is (Hochlaf or Gelfand).
Although the crossing position and corresponding SOC values may be different, we do not think it would significantly change the overall cross section (and rate coefficients) \red{because of the relatively small contribution of the SOC pathways of $\mathrm{^3\Pi_u}$ states to overall \ch{N(^2P) -> N(^2D)} cross sections (see Fig.~\ref{fig:compare_PEC}b)}.

Therefore, we believe that the mixed usage of PEC results from the two papers should not substantially change the results and therefore we are generating high quality results. \red{Moreover, our comparison shows that for moderate to low energy dynamics, the long range and asymptotic behavior of the PECs may be more relevant, compared to the high energy part.} Nevertheless, in future work, we look forward to a more accurate calculation with both spin-orbit couplings and nonadiabatic couplings evaluated on the same footing \red{for a verification}.

\section{Conclusion} \label{sec:conclusion}

This paper has provided a comprehensive description of the inelastic scattering dynamics associated with two nitrogen atoms in states that include \ch{N(^4S)}, \ch{N(^2D)} and \ch{N(^2P)}, using accurate PECs and couplings that including both SOCs and nonadiabatic (derivative) couplings, including all partial waves that enable the calculations of cross sections and rate coefficients over a wide range of energies and temperatures of relevance to hypersonic flight in the upper atmosphere. Other than the PEC's and couplings, the only approximations in the calculation involve assuming that the rotational angular momentum is fixed during collisions, that transitions involving SOCs are perturbative and that one can include for contributions from different fine structure transitions with a perturbation theory approximation. For the process \ch{N(^4S) + N <-> N(^2P/^2D) + N}, we find that the cross section is dominated by a single transition that is governed by SOC, leading to small cross sections for both excitation and quenching. Also the calculated rate coefficients for the \ch{N(^2D)} final state and its reverse are similar to that recently calculated for the corresponding process \ch{N(^4S) + N_2 <-> N(^2D) + N_2}. 
For the \ch{N(^2D) <-> N(^2P)} process, we find pathways for the nonadiabatic dynamics which have important temperature-dependence: At low temperature the collisions are dominated by intersystem crossings, but at high temperature (> 10000 K), nonadiabatic transitions become dominant. Our predicted rate coefficient for quenching is consistent with two of the three experimental results, so we recommend using the results which agree for modeling applications. 

We have further examined the processes \ch{N + N(^2P) -> N(^2D) + N(^2D)} and \ch{N + N(^2P) <-> N + N(^4S)}. We suggest that the former process could be an important pathway for \ch{N(^2P)} quenching at extremely high temperature (> 15000 K). The latter process is also activated in the high temperature regime, however it may not be competitive with the \ch{N(^2P) -> N(^2D)} pathway according to our analysis. Future studies should address these two processes for clarification.

\section*{Acknowledgements}
This work is supported by the Office of Naval Research under MURI grant N00014-22-1-2661. We also thank Igor Adamovich, Natalia Gelfand and Iain Boyd for helpful discussions.

%aipnum4-2.bst 2019-01-14 (MD) hand-edited version of apsrev4-1.bst

\end{document}